\documentclass[amsmath,aps,showpacs,a4paper,10pt]{revtex4}

 \usepackage{epsf}
 \usepackage{graphicx}    

\begin{document}

 \newcommand{\be}[1]{\begin{equation}\label{#1}}
 \newcommand{\ee}{\end{equation}}
 \newcommand{\bea}{\begin{eqnarray}}
 \newcommand{\eea}{\end{eqnarray}}
 \def\disp{\displaystyle}

 \def\gsim{ \lower .75ex \hbox{$\sim$} \llap{\raise .27ex \hbox{$>$}} }
 \def\lsim{ \lower .75ex \hbox{$\sim$} \llap{\raise .27ex \hbox{$<$}} }

 \begin{titlepage}

 \begin{flushright}
 arXiv:1707.06367
 \end{flushright}

 \title{\Large \bf Model-Independent Constraints
 on Lorentz Invariance Violation via the Cosmographic Approach}

 \author{Xiao-Bo~Zou\,}
 \email[\,email address:\ ]{1258646432@qq.com}
 \affiliation{School of Physics,
 Beijing Institute of Technology, Beijing 100081, China}

 \author{Hua-Kai~Deng\,}
 \affiliation{School of Physics,
 Beijing Institute of Technology, Beijing 100081, China}

 \author{Zhao-Yu~Yin\,}
 \affiliation{School of Physics,
 Beijing Institute of Technology, Beijing 100081, China}

 \author{Hao~Wei\,}
 \email[\,Corresponding author;\ email address:\ ]{haowei@bit.edu.cn}
 \affiliation{School of Physics,
 Beijing Institute of Technology, Beijing 100081, China}

 \begin{abstract}\vspace{1cm}
 \centerline{\bf ABSTRACT}\vspace{2mm}
 Since Lorentz invariance plays an important role in modern
 physics, it is of interest to test the possible Lorentz
 invariance violation (LIV). The time-lag (the arrival time
 delay between light curves in different energy bands) of
 Gamma-ray bursts (GRBs) has been extensively used to this
 end. However, to our best knowledge, one or more particular
 cosmological models were assumed {\it a~priori} in (almost)
 all of the relevant works in the literature. So, this makes
 the results on LIV in those works model-dependent and hence
 not so robust in fact. In the present work, we try to avoid
 this problem by using a model-independent approach. We
 calculate the time delay induced by LIV with the cosmic
 expansion history given in terms of cosmography, without
 assuming any particular cosmological model. Then, we constrain
 the possible LIV with the observational data, and find weak
 hints for LIV.
 \end{abstract}


 \maketitle

 \end{titlepage}

 \renewcommand{\baselinestretch}{1.0}


\section{Introduction}\label{sec1}

As is well known, Lorentz invariance plays an important role
 in modern physics. Actually, it is one of the foundation
 stones of special/general relativity and particle physics,
 which have been well tested in solar system and colliders.
 If Lorentz invariance is violated, the pillars of modern
 physics will be shocked and new physics is needed. So, it is
 of interest to test the possible Lorentz invariance violation
 (LIV) with various terrestrial experiments
 and astrophysical/cosmological
 observations~\cite{Mattingly:2005re,Liberati:2013xla}.

In the literature, there exist many theories inducing LIV. Here
 we are interested in the possible violation of Lorentz
 invariance induced by quantum gravity (QG). Commonly, most
 theories of QG (e.g. string theory, loop quantum gravity,
 doubly special relativity) predict that LIV might happen on
 high energy scales~\cite{AmelinoCamelia:1997gz,Ellis:2002in,
 Ellis:2005wr,Jacob:2008bw,Ellis:2011ek,Ackermann:2009aa,
 Biesiada:2009zz,Chang:2015qpa,Xu:2016zxi,Xu:2016zsa,Nemiroff:2011fk,
 Vasileiou:2013vra,Wei:2016exb,Pan:2015cqa,Guo:2012mv,
 Mewesworks}. The propagation of high energy photons through
 the spacetime foam might exhibit a non-trivial dispersion relation
 in vacuum (which should be regarded as a non-trivial medium in QG).
 The deformed dispersion relation for photons usually takes the
 form $p^2 c^2=E^2 [\,1 + f(E/E_{EQ})]$, where $E_{QG}$ is the
 effective QG energy scale, $f$ is a dimensionless function
 depending on the particular QG model, $c$ is the limiting
 speed of light on low energy scales, $p$ and $E$ are the
 momentum and energy of photons, respectively. On low energy
 scales $E\ll E_{QG}$, one can consider a series expansion of
 this dispersion relation, namely $p^2 c^2=E^2[\,1+\xi E/E_{QG}
 +{\cal O}(E^2/E_{QG}^2)]$, where $\xi=\pm 1$ is a sign
 ambiguity~\cite{AmelinoCamelia:1997gz}. Such a series
 expansion corresponds to an energy-dependent speed of light,
 $v=\partial E/\partial p\sim c\left(1-\xi
 E/E_{QG}\right)$~\cite{AmelinoCamelia:1997gz}. So, the
 high and low energy photons will not reach us at the same
 time. A signal of energy $E$ that travels a distance $L$
 acquires a time delay (measured with respect to the ordinary
 case of an energy-independent speed $c$), namely
 $\Delta t\sim\xi (E/E_{QG})(L/c)$. Although the QG effect is
 expected to be very weak (since $E_{QG}$ is typically close
 to the Planck energy scale $E_{P}\sim 10^{19}\,{\rm GeV}$),
 a very long distance $L$ can still make it testable. In the
 pioneer work~\cite{AmelinoCamelia:1997gz}, Amelino-Camelia
 {\it et al.} proposed that Gamma-ray bursts (GRBs) at a
 cosmological distance can be used to test the possible
 LIV, while time-lag (the arrival time delay between light
 curves in different energy bands) is a common feature in
 GRBs~\cite{timelag}.

Following the pioneer work of Amelino-Camelia
 {\it et al.}~\cite{AmelinoCamelia:1997gz}, many constraints
 on LIV have been obtained from the time-lag of GRBs in the
 literature (e.g.~\cite{Ellis:2005wr,Jacob:2008bw,Ellis:2011ek,
 Ackermann:2009aa,Biesiada:2009zz,Chang:2015qpa,Xu:2016zxi,
 Xu:2016zsa,Nemiroff:2011fk,Vasileiou:2013vra,Wei:2016exb,
 Pan:2015cqa}). It is worth noting that the cosmic expansion
 history (characterized by the Hubble parameter $H(z)$ as a
 function of redshift $z$) should be taken into account when
 one calculates the time delay $\Delta t$ (see Sec.~\ref{sec2}
 for details), since photons propagate through the expanding
 space. Thus, the cosmic expansion history $H(z)$ should be
 given in advance. Actually, a particular cosmological model
 was assumed {\it a priori} in (almost) all of the relevant
 works. For example, the well-known spatially flat $\Lambda$CDM
 model $H(z)=H_0\sqrt{\Omega_M(1+z)^3+\Omega_\Lambda}$ is
 usually assumed in the literature (e.g.~\cite{Ellis:2005wr,
 Jacob:2008bw,Ellis:2011ek,Ackermann:2009aa,Biesiada:2009zz,
 Chang:2015qpa,Xu:2016zxi,Xu:2016zsa,Nemiroff:2011fk,Vasileiou:2013vra,
 Wei:2016exb}), while all the values of the model parameters
 $\Omega_M$, $\Omega_\Lambda=1-\Omega_M$, and the Hubble
 constant $H_0$ are fixed (typically taken from the WMAP/Planck
 results). In a few of works, some other cosmological models
 instead of $\Lambda$CDM model are also considered. For
 instance, in e.g.~\cite{Biesiada:2009zz}, the dark energy
 models with $w=const.$ and $w=w_0+w_1 z$, the generalized
 Chaplygin gas (GCG) model, and the Dvali-Gabadadze-Porrati
 (DGP) braneworld model are assumed, while all the values of
 the corresponding model parameters have also been fixed.
 Different from the above works, in~\cite{Pan:2015cqa}, the
 corresponding model parameters are not fixed {\it a priori},
 and they are constrained together with the LIV parameters by
 using the observational data. However, some particular
 cosmological models still should be assumed in~\cite{Pan:2015cqa},
 i.e. the flat $\Lambda$CDM model, the $w$CDM model, and
 the Chevallier-Polarski-Linder~(CPL) model, although their
 model parameters are free. In summary, to our best knowledge,
 one or more particular cosmological models have been assumed
 {\it a~priori} in (almost) all of the relevant works in the
 literature. So, this makes the results on LIV in those
 works model-dependent and hence not so robust in fact.

In the present work, we try to avoid this problem by using a
 model-independent approach. As is well known, one of the
 powerful model-independent approaches is the so-called
 cosmography~\cite{Weinberg1972,Visser:2003vq,Bamba:2012cp,
 Cattoen:2008th,Vitagliano:2009et,Cattoen:2007id,Xu:2010hq,
 Xia:2011iv,Zhang:2016urt,Dunsby:2015ers,Zhou:2016nik,Luongoworks}.
 In fact, the only necessary assumption of cosmography is the
 cosmological principle. With cosmography, one can analyze the
 evolution of the universe without assuming any underlying
 theoretical model. Essentially, cosmography is the Taylor
 series expansion of the quantities related to the cosmic expansion
 history, such as the scale factor $a(t)$, the Hubble parameter
 $H(z)$ and the luminosity distance $d_L(z)$. Therefore, this
 makes cosmography model-independent indeed. We refer to
 e.g.~\cite{Weinberg1972,Visser:2003vq,Bamba:2012cp,Cattoen:2008th,
 Vitagliano:2009et,Cattoen:2007id,Xu:2010hq,Xia:2011iv,
 Zhang:2016urt,Dunsby:2015ers,Zhou:2016nik,Luongoworks} and
 references therein for more details of cosmography. So, in the
 present work, we can calculate the time delay $\Delta t$
 induced by LIV with the cosmic expansion history given in terms of
 cosmography, without assuming any particular cosmological
 model, unlike the relevant works on LIV mentioned above. The
 results on LIV obtained via the cosmographic approach will
 be model-independent and robust.

The rest of this paper is organized as follow. In
 Sec.~\ref{sec2}, we briefly review the formalism of the time
 delay $\Delta t$ induced by LIV. In Sec.~\ref{sec3}, we derive
 the cosmic expansion history in terms of cosmography with
 respect to redshift $z$ at first. Then, we constrain the LIV
 parameters together with the cosmographic parameters by using
 the time delay data from GRBs, the observational data from
 type Ia supernovae (SNIa) and the baryon acoustic oscillation
 (BAO). Note that we adopt the Markov Chain Monte Carlo (MCMC)
 technique in doing this. As is well known, there exists a
 divergence problem in cosmography with respect to redshift
 $z$ when $z>1$. Thus, another cosmography with respect to the
 so-called $y$-shift $y\equiv z/(1+z)$ has been proposed
 in the literature, which alleviates the divergence problem
 since $y<1$ in the range of $0\leq z<\infty$. In
 Sec.~\ref{sec4}, we obtain the observational constraints on
 LIV via the cosmographic approach with respect to $y$-shift.
 The brief concluding remarks are given in Sec.~\ref{sec5}.


\section{Time delay of GRB photons induced by LIV}\label{sec2}

As mentioned in Sec.~\ref{sec1}, the deformed
 dispersion relation for photons~\cite{AmelinoCamelia:1997gz}
 usually takes the form $p^2 c^2=E^2 [\,1 + f(E/E_{EQ})]$,
 where $E_{QG}$ is the effective QG energy scale, $f$ is a
 dimensionless function depending on the particular QG model,
 $c$ is the limiting speed of light on low energy scales, $p$
 and $E$ are the momentum and energy of photons, respectively.
 On low energy scales $E\ll E_{QG}$, one can always expand this
 deformed dispersion relation as a Taylor series
 (see e.g.~\cite{Vasileiou:2013vra}),
 \be{eq1}
 E^2=p^2 c^2\left[1-\sum\limits_{n=1}^\infty s_\pm\left(
 \frac{E}{\xi_n E_{QG}}\right)^n\right]\,,
 \ee
 where $s_\pm=\pm 1$ is the ``sign of LIV'', a theory-dependent
 factor equal to $+1$ ($-1$) for a decrease (increase)
 in photon speed with an increasing photon
 energy~\cite{Vasileiou:2013vra}. $\xi_n$ is a dimensionless
 parameter, and $E_{QG,n}=\xi_n E_{QG}$ is actually the
 effective energy scale where LIV happens for the order $n$
 term~\cite{Vasileiou:2013vra}. For $E\ll E_{QG}$, the lowest
 order term in the series not suppressed by theory (usually
 the $n=1$ term) is expected to dominate the sum. If the $n=1$
 term is suppressed (say, by a symmetry law), the next term
 $n=2$ will dominate, and so forth. If the dominated LIV
 correction is of order $n$, Eq.~(\ref{eq1}) can be approximated by
 \be{eq2}
 E^2=p^2 c^2\left[1-s_\pm\left(
 \frac{E}{\xi_n E_{QG}}\right)^n\,\right]\,,
 \ee
 which is the well-known form adopted in most of the relevant
 works in the literature (e.g.~\cite{Ellis:2005wr,Jacob:2008bw,
 Ellis:2011ek,Ackermann:2009aa,Biesiada:2009zz,Chang:2015qpa,
 Xu:2016zxi,Xu:2016zsa,Nemiroff:2011fk,Vasileiou:2013vra,
 Wei:2016exb,Pan:2015cqa}). Note that $E$ in the right hand
 side of Eq.~(\ref{eq2}) can be freely substituted with
 $pc$ since we are only interested in the leading order
 correction~\cite{Jacob:2008bw}. Keeping this in mind and using
 Eq.~(\ref{eq2}), the energy-dependent speed of photons is
 given by (see e.g.~\cite{Vasileiou:2013vra,
 Xu:2016zxi,Xu:2016zsa,Wei:2016exb})
 \be{eq3}
 v=\frac{\partial E}{\partial p}=c\left[1-
 s_\pm\frac{n+1}{2}\left(\frac{E}{\xi_n E_{QG}}\right)^n\,\right]\,.
 \ee
 Following the standard procedures given in e.g.~\cite{Jacob:2008bw}
 (calculating the comoving path in the expanding universe is
 the key), one can finally get the LIV-induced time delay
 between photons with energies $E_{\rm high}$ and $E_{\rm low}$
 as (see e.g.~\cite{Xu:2016zxi,Xu:2016zsa,Vasileiou:2013vra})
 \be{eq4}
 \Delta t_{LIV}=s_\pm\frac{1+n}{2H_0}\frac{E_{\rm high}^n-
 E_{\rm low}^n}{E_{QG,n}^n}\int_0^z\frac{(1+
 \tilde{z})^n d\tilde{z}}{h(\tilde{z})}\,,
 \ee
 where $h(z)\equiv H(z)/H_0$ is the dimensionless Hubble parameter,
 and $z$ is the redshift of GRB. Following most of the relevant
 works in the literature (e.g.~\cite{Ellis:2005wr,Jacob:2008bw,
 Ellis:2011ek,Ackermann:2009aa,Biesiada:2009zz,Chang:2015qpa,
 Xu:2016zxi,Xu:2016zsa,Nemiroff:2011fk,Vasileiou:2013vra,
 Pan:2015cqa}), we only consider the case of $n=1$, $s_\pm=+1$
 and $\xi_1=1$ in the present work. In this case,
 Eq.~(\ref{eq4}) becomes
 \be{eq5}
 \Delta t_{LIV}=\frac{\Delta E}{H_0 E_{QG}}
 \int_0^z\frac{\left(1+\tilde{z}\right)d\tilde{z}}{h(\tilde{z})}\,.
 \ee
 Note that $\Delta t_{LIV}$ in the original published version
 of~\cite{Ellis:2005wr} lacked the factor $(1+z)$ in the
 integration, and it has been corrected in the Erratum while
 the conclusion was modified accordingly.

For a cosmic transient source (e.g. GRB), the observed time
 delay between two different energy bands should include
 five terms~\cite{Wei:2015hwd,Gao:2015lca},
 \be{eq6}
 \Delta t_{obs}=\Delta t_{LIV}+\Delta t_{\rm int}
 +\Delta t_{\rm spe}+\Delta t_{\rm DM}+\Delta t_{\rm gra}\,,
 \ee
 where $\Delta t_{LIV}$ is the LIV-induced time delay as
 mentioned above. $\Delta t_{\rm int}$ is the intrinsic
 (astrophysical) time delay, which means that photons with
 high and low energies do not leave the source simultaneously.
 This term is difficult to predict because we have no good
 understanding on the physics of source evolution. As in the
 literature (e.g.~\cite{Ellis:2005wr,Biesiada:2009zz,Pan:2015cqa}),
 one can write it as $\Delta t_{\rm int}=b\left(1+z\right)$
 while the cosmic expansion has been taken into account, and
 the constant parameter $b$ characterizes our ignorance.
 $\Delta t_{\rm spe}$ represents the potential time delay due
 to special relativistic effects if photons have a non-zero
 rest mass. Since modern experiments have provided the upper
 limits for the photon rest mass as
 $m_{ph}<10^{-18}\,{\rm eV}/c^2$~\cite{Amsler:2008zzb}, this
 term is negligible in fact~\cite{Gao:2015lca}. $\Delta t_{\rm DM}$
 is the time delay contribution from the dispersion by the
 line-of-sight free electron content, which is also negligible
 for GRB photons~\cite{Wei:2015hwd}. $\Delta t_{\rm gra}$
 represents the effect of gravitational potential along the
 propagation path of photons if the Einstein's equivalence
 principle (EEP) is violated. This term can be dropped since
 EEP is preserved in our case. Substituting Eq.~(\ref{eq5})
 and $\Delta t_{\rm int}=b\left(1+z\right)$ into
 Eq.~(\ref{eq6}), and neglecting other terms, we
 obtain~\cite{Ellis:2005wr,Ellis:2011ek,Biesiada:2009zz,
 Xu:2016zxi,Xu:2016zsa,Pan:2015cqa}
 \be{eq7}
 \frac{\Delta t_{obs}}{1+z}=a_{LIV}K+b\,,
 \ee
 where $a_{LIV}\equiv\Delta E/(H_0 E_{QG})$, and
 \be{eq8}
 K\equiv\frac{1}{1+z}
 \int_0^z\frac{(1+\tilde{z})\,d\tilde{z}}{h(\tilde{z})}\,.
 \ee
 Obviously, if $a_{LIV}=0$, there is no LIV. On the other hand,
 if the evidence of $a_{LIV}\not=0$ is found, LIV happens on
 the energy scales above $E_{QG}$.


\section{Constraints on LIV via the cosmographic approach
 with~respect~to redshift $z$}\label{sec3}


\subsection{Cosmographic approach with respect
 to redshift $z$}\label{sec3a}

For convenience, we recast Eq.~(\ref{eq7}) as
 \be{eq9}
 \Delta t_{obs} = a_{LIV}{\cal K} + b\left(1+z\right)\,,
 \ee
 where
 \be{eq10}
 {\cal K}\equiv\left(1+z\right)K=
 \int_0^z\frac{\left(1+\tilde{z}\right)d\tilde{z}}{h(\tilde{z})}\,.
 \ee
 In order to constrain the possible LIV, one should calculate
 the theoretical time delay induced by LIV,
 $\Delta t_{th}=a_{LIV}{\cal K}+b\left(1+z\right)$, and then
 confront it with the observed one, $\Delta t_{obs}$. From
 Eq.~(\ref{eq10}), it is easy to see that the cosmic expansion
 history (characterized by $h(z)$) should be given in advance.
 As mentioned in Sec.~\ref{sec1}, we will present it via the
 cosmographic approach which is model-independent, rather than
 assuming a particular cosmological model (e.g. $\Lambda$CDM)
 as in the literature.

The only necessary assumption of cosmography is
 the cosmological principle, so that the spacetime metric is
 the one of the Friedmann-Robertson-Walker (FRW) universe,
 \be{eq11}
 ds^2=-c^2 dt^2+a^2(t)\left[\frac{dr^2}{1-kr^2}+
 r^2\left(d\theta^2+\sin^2 \theta\,d\phi^2\right)\right]\,,
 \ee
 in terms of the comoving coordinates $(t,\,r,\,\theta,
 \,\phi)$, where $a$ is the scale factor. Motivated by the
 inflation paradigm and the observational results from e.g.
 Planck 2015 data~\cite{Ade:2015xua}, in this work we only
 consider a spatially flat FRW universe with $k=0$. Introducing
 the so-called cosmographic parameters, namely the Hubble
 constant $H_0$, the deceleration $q_0$, the jerk $j_0$, the
 snap $s_0$,
 \be{eq12}
 H_0\equiv\left.\frac{1}{a}\frac{da}{dt}\right|_{t=t_0}\,,\quad
 q_0\equiv\left.-\frac{1}{aH^2}\frac{d^{2}a}{dt^{2}}\right|_{t=t_0}\,,\quad
 j_0\equiv\left.\frac{1}{aH^3}\frac{d^{3}a}{dt^{3}}\right|_{t=t_0}\,,\quad
 s_0\equiv\left.\frac{1}{aH^4}\frac{d^{4}a}{dt^{4}}\right|_{t=t_0}\,,\quad
 \ee
 one can expand the scale factor $a$ in terms of a Taylor
 series with respect to cosmic
 time $t$~\cite{Weinberg1972,Visser:2003vq},
 \be{13}
 a(t)=a(t_0)\left[1 + H_0 (t-t_0) - \frac{q_0}{2}
 H_{0}^{2}(t-t_{0})^{2}+\frac{j_{0}}{3!} H_{0}^{3}(t-t_{0})^{3}
 +\frac{s_{0}}{4!} H_{0}^{4} (t-t_{0})^{4}
 +{\cal O}\left((t-t_{0})^5\right)\right]\,.
 \ee
 One of the most important quantities in cosmology is the
 luminosity distance $d_L=\left(c/H_0\right)D_L$, where the
 dimensionless luminosity distance $D_L$ is
 defined by~\cite{Weinberg1972}
 \be{eq14}
 D_L\equiv(1+z)\int_0^z\frac{d\tilde{z}}{h(\tilde{z})}\,.
 \ee
 We can also expand the dimensionless luminosity distance $D_L$
 (equivalent to $d_L$) in terms of a Taylor series with respect
 to redshift $z$ (see e.g.~\cite{Bamba:2012cp,Weinberg1972,
 Visser:2003vq,Zhou:2016nik} for details),
 \bea
 D_L(z)&=&z+\frac{1}{2}\left(1-q_0\right)z^2
 -\frac{1}{6}\left(1-q_0-3q_0^2+j_0\right)z^3\nonumber\\[1mm]
 &&+\frac{1}{24}\left(2-2q_0-15q_0^2-15q_0^3+5j_0+10q_0 j_0
 +s_0\right)z^4+{\cal O}\left(z^5\right)\,.\label{eq15}
 \eea
 Differentiating Eq.~(\ref{eq14}), we get
 \be{eq16}
 \frac{1+z}{h(z)}=\frac{d D_L}{dz}-\frac{D_L}{1+z}\,.
 \ee
 Substituting Eqs.~(\ref{eq16}) and (\ref{eq15})
 into Eq.~(\ref{eq10}), we obtain
 \bea
 {\cal K}(z) &=& z-\frac{q_0}{2}z^2+\left(\frac{q_0}{3}
 +\frac{q_0^2}{2}-\frac{j_0}{6}\right)z^3\nonumber\\[1mm]
 & & +\left(-\frac{q_0}{4}-\frac{3}{4}q_0^2-\frac{5}{8}q_0^3
 +\frac{j_0}{4}+\frac{5}{12} q_0 j_0+\frac{s_0}{24}\right)z^4
 +{\cal O}\left(z^5\right)\,.\label{eq17}
 \eea
 In this approach, one can calculate the theoretical time delay
 $\Delta t_{th}=a_{LIV}{\cal K}+b\left(1+z\right)$ by using
 Eq.~(\ref{eq17}), without assuming any particular cosmological
 model. The LIV parameters $a_{LIV}$ and $b$, together with
 the cosmographic parameters, will be determined by using the
 observational data. Note that one can also obtain $h(z)$ or
 $1/h(z)$ by using Eq.~(\ref{eq16}) with $D_L$ given in
 Eq.~(\ref{eq15}). This is useful when one calculates other
 quantities (e.g. $D_V(z)$ used below).


\subsection{Observational data}\label{sec3b}

Ellis {\it et al.}~\cite{Ellis:2002in} have developed the
 systematic analysis of statistical samples of GRBs at a range
 of different redshifts, and they have introduced techniques
 from signal processing such as wavelet analysis to identify
 and correlate genuine features in the intensities observed in
 different energy bands. Later, using these techniques, Ellis
 {\it et al.}~\cite{Ellis:2005wr} compiled a time delay dataset
 from 35 GRBs with known redshifts from $z=0.168$ to $z=6.29$.
 The numerical dataset can be found in Table 1
 of~\cite{Ellis:2005wr}. To constrain the possible LIV, we
 perform the $\chi^2$ statistics. The $\chi^2$ from the
 time-lags of GRBs is given by
 \be{eq18}
 \chi^2_{GRB}=\sum\limits_{i=1}^{35}\left[
 \frac{\Delta t_{th}(z_i)-\Delta t_{obs}(z_i)}{\sigma_i}\right]^2\,,
 \ee
 where $\Delta t_{obs}$ and $\sigma_i$ are the observed time
 delay and the corresponding uncertainty given in Table 1
 of~\cite{Ellis:2005wr}, and
 $\Delta t_{th}=a_{LIV}{\cal K}+b\left(1+z\right)$ is the
 theoretical time delay, while $\cal K$ is given
 in Eq.~(\ref{eq17}).

Clearly, the cosmographic parameters (characterizing the cosmic
 expansion history) cannot be well constrained by using only
 the time delay data from GRBs. So, the other observational data are
 needed. Here, we consider the JLA (joint light-curve analysis)
 dataset~\cite{Betoule:2014frx} consisting of 740 SNIa obtained
 by the SDSS-II and SNLS collaborations. The theoretical
 distance modulus is defined by~\cite{Weinberg1972,Betoule:2014frx}
 \be{eq19}
 \mu_{th}=5\log_{10}\frac{d_L}{\rm Mpc}+25\,,
 \ee
 where the luminosity distance $d_L=\left(c/H_0\right)D_L$, and
 $D_L$ is given in Eq.~(\ref{eq15}). On the other hand, in the
 JLA dataset, the observed distance modulus is
 given by~\cite{Betoule:2014frx}
 \be{eq20}
 \mu_{obs}=m_B-\left({\cal M}-
 \alpha X_1+\beta{\cal C}\right)\,,
 \ee
 where $m_B$ corresponds to the observed peak magnitude in
 rest frame B band. $\alpha$ and $\beta$ are both nuisance
 parameters. $X_1$ and $\cal C$ are the stretch measure and the
 color measure of SNIa, respectively. $\cal M$ is a nuisance
 parameter representing some combination of the absolute
 magnitude of a fiducial SNIa and the Hubble constant $H_0$.
 The $\chi^2$ from JLA SNIa reads~\cite{Betoule:2014frx}
 \be{eq21}
 \chi^2_{JLA}=\Delta\vec{\mu}^{\; T}\,{\bf C}^{-1}\Delta\vec{\mu}\,,
 \ee
 where $\Delta\mu=\mu_{obs}-\mu_{th}$, and $\bf C$ is the covariance
 matrix of $\vec{\mu}$. It is equivalent to~\cite{Conley:2011ku} (see
 also e.g.~\cite{Wang:2015tua})
 \be{eq22}
 \chi^2_{JLA}=\Delta\vec{m}^{\,T}\,{\bf C}^{-1}
 \Delta\vec{m}\,,
 \ee
 where $\Delta m = m_B - m_{\rm mod}$, and
 \be{eq23}
 m_{\rm mod}=5\log_{10}D_L-\alpha X_1+\beta{\cal C}+{\cal M}\,,
 \ee
 while $H_0$ in $d_L$ can be absorbed into $\cal M$. The
 numerical data of $m_B$, $X_1$, $\cal C$, and the covariance
 matrix $\bf C$ can be found from the JLA
 dataset~\cite{Betoule:2014frx,JLAreadme}. The nuisance
 parameters $\cal M$, $\alpha$ and $\beta$ could
 be marginalized (note that $H_0$ is also a nuisance parameter
 and it can be absorbed into $\cal M$, and hence the JLA SNIa
 dataset is Hubble-free in fact). It is worth noting that
 since March 2014, the JLA likelihood plugin is included
 in the official release of the MCMC code
 CosmoMC~\cite{Lewis:2002ah,CosmoMCsite}.

We can further consider the observational data from BAO. Here
 we use the data $D_V(0.35)/D_V(0.2)=1.736 \pm 0.065$ from
 SDSS Collaboration~\cite{Percival:2009xn}, where $D_V(z)$ is
 related to the angular diameter distance $d_A$ and the
 luminosity distance $d_L$ or $D_L$ according
 to~\cite{Eisenstein:2005su,Percival:2009xn}
 \be{eq24}
 D_V(z)\equiv\left[(1+z)^2 d_A^2 \frac{cz}{H(z)}\right]^{1/3}
 =\left[\frac{d_L^2}{(1+z)^2}\frac{cz}{H(z)}\right]^{1/3}
 =\frac{c}{H_0}\left[\frac{D_L^2}{(1+z)^2}\frac{z}{h(z)}
 \right]^{1/3}\,,
 \ee
 in which we have used the well-known relation
 $d_A=d_L/(1+z)^2$ (see e.g.~\cite{Weinberg1972,
 Cattoen:2008th,Cattoen:2007id}) and
 $d_L=\left(c/H_0\right)D_L$. Note that the factor $c/H_0$ will
 be cancelled in $D_V(0.35)/D_V(0.2)$ and hence it is also
 Hubble-free. One can calculate $D_V(z)$ by using $D_L$ given
 in Eq.~(\ref{eq15}),
 and $z/h(z)=[\,z/(1+z)\,]\cdot [\,dD_L/dz-D_L/(1+z)\,]$
 from Eq.~(\ref{eq16}). The $\chi^2$ from BAO is given by
 \be{eq25}
 \chi^2_{BAO}=\left[
 \frac{D_{V}(0.35)/D_{V}(0.2)-1.736}{0.065}\right]^2\,.
 \ee
 Note that actually there exist other BAO data in
 the literature, such as the observational data
 of $A\equiv\Omega_{m0}^{1/2}H_0
 D_V(z)/z$~\cite{Eisenstein:2005su,Blake:2011en,Wei:2017mzf},
 and $d_z\equiv r_s(z_d)/D_V(z)$~\cite{Percival:2009xn,
 Beutler:2011hx,Cai:2014ela}. However, they will introduce
 one or more free model parameters such as $\Omega_{m0}$,
 $\Omega_b$, while the Hubble constant $H_0$ is no longer
 a nuisance parameter. This is a drawback, and makes the
 constraints loose. So, we do not consider such types of
 BAO data in this work. Similarly, we also do not consider the
 observational data from cosmic microwave background (CMB),
 since one or more free model parameters, e.g. $\Omega_{m0}$,
 and/or $\Omega_b$, $H_0$, should be introduced.


 \begin{center}
 \begin{figure}[tbp]
 \centering
 \vspace{-12mm}  
 \includegraphics[width=1.0\textwidth]{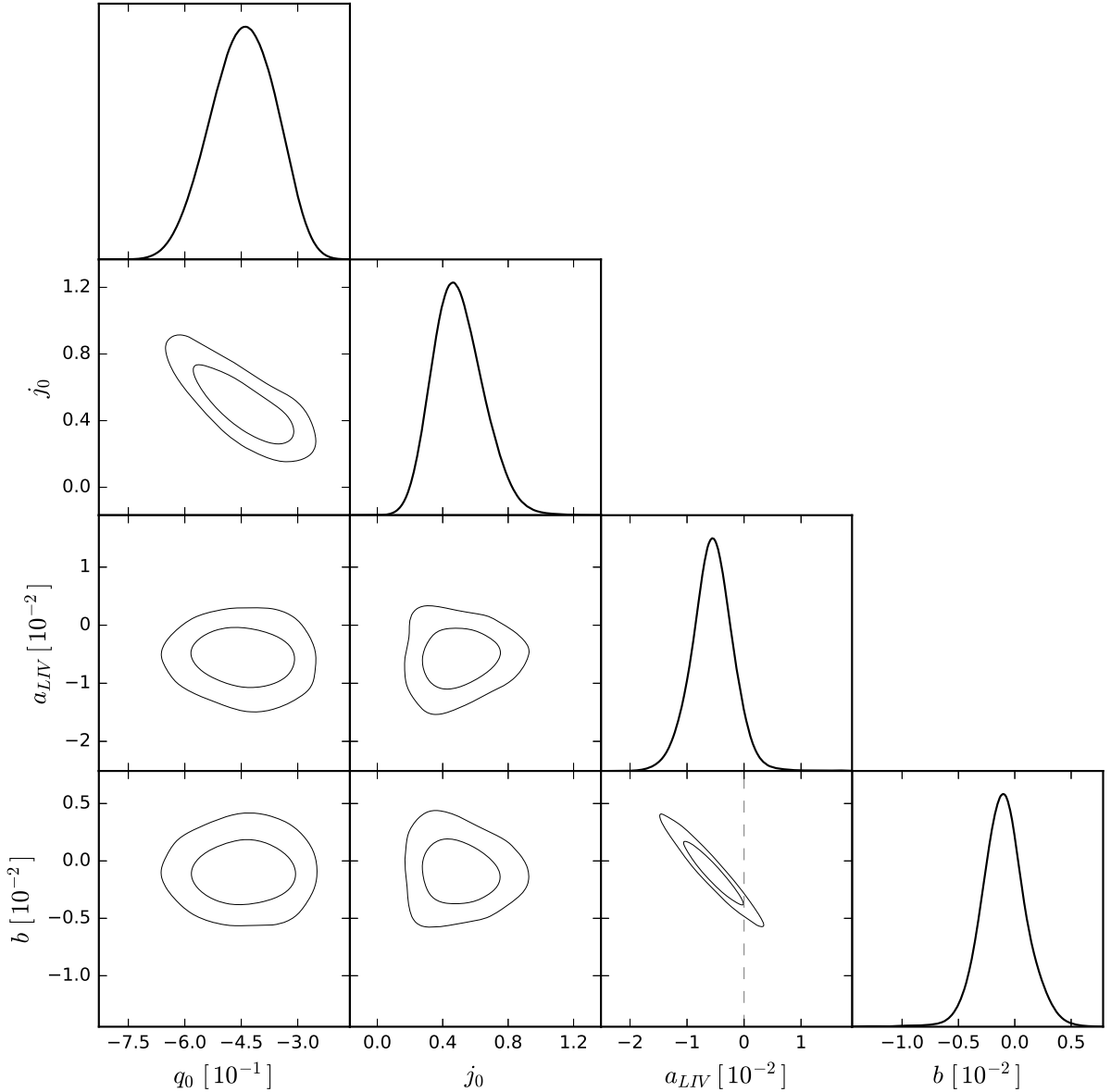}
 \caption{\label{fig1} The 2D marginalized $1\sigma$ and
 $2\sigma$ contours and the 1D distributions of the LIV
 parameters and the cosmographic parameters for the case of 3rd
 order cosmography with respect to redshift $z$ (labeled as
 ``\,$z-j$\,''). Note that $a_{LIV}=0$ is also indicated by a
 dashed line in the $a_{LIV}-b$ panel, and $q_0$, $a_{LIV}$,
 $b$ are given in units of $10^{-1}$, $10^{-2}$, $10^{-2}$,
 respectively. See the text and Table~\ref{tab1} for details.}
 \end{figure}
 \end{center}



 \begin{center}
 \begin{figure}[tbp]
 \centering
 \vspace{-12mm}  
 \includegraphics[width=1.0\textwidth]{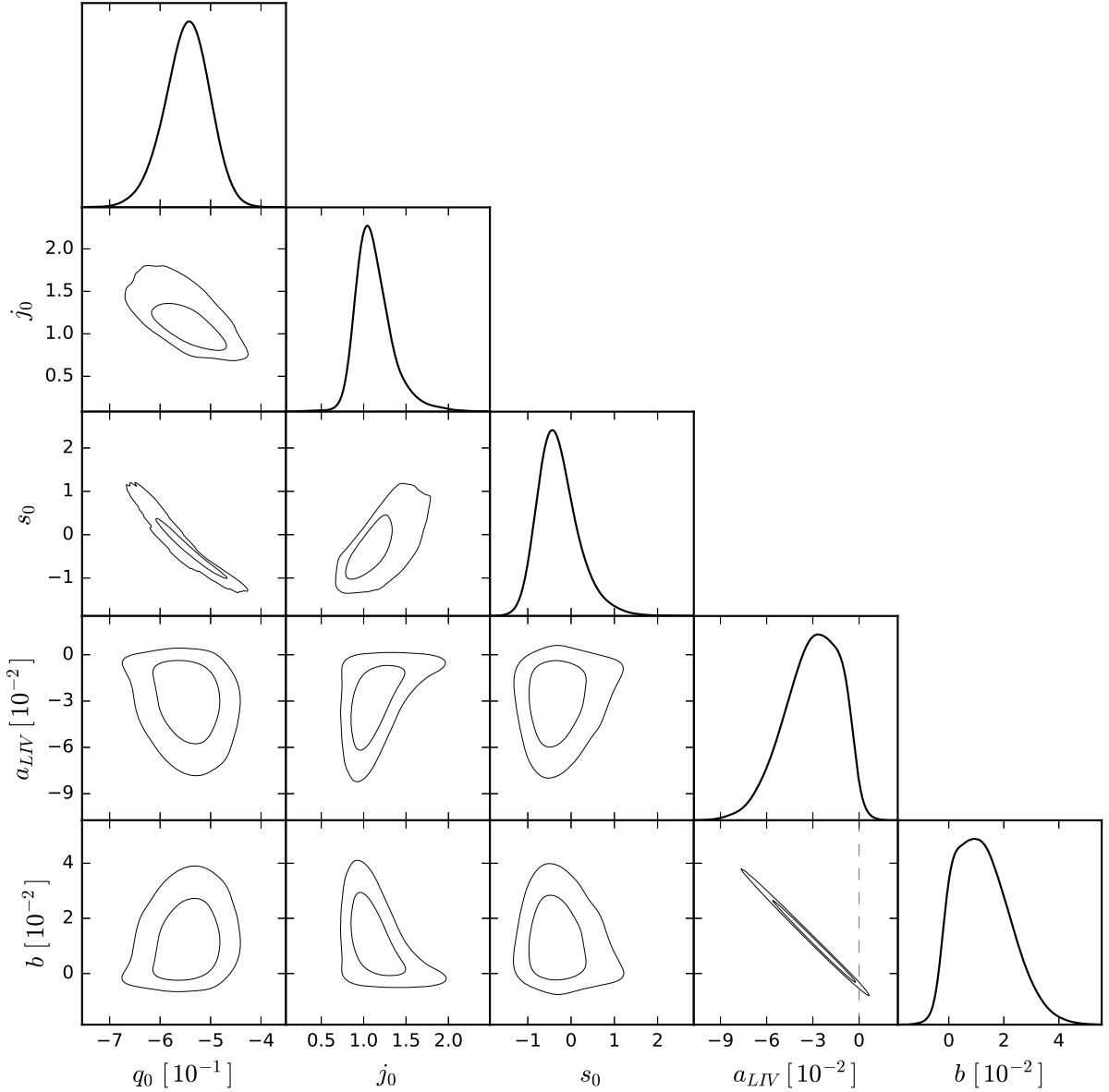}
 \caption{\label{fig2} The 2D marginalized $1\sigma$ and
 $2\sigma$ contours and the 1D distributions of the LIV
 parameters and the cosmographic parameters for the case of 4th
 order cosmography with respect to redshift $z$ (labeled as
 ``\,$z-s$\,''). Note that $a_{LIV}=0$ is also indicated by a
 dashed line in the $a_{LIV}-b$ panel, and $q_0$, $a_{LIV}$,
 $b$ are given in units of $10^{-1}$, $10^{-2}$, $10^{-2}$,
 respectively. See the text and Table~\ref{tab1} for details.}
 \end{figure}
 \end{center}


\vspace{-25mm}  


\subsection{Constraints on LIV}\label{sec3c}

Here, we constrain the LIV parameters $a_{LIV}$ and $b$,
 together with the cosmographic parameters, by using the
 observational data mentioned in Sec.~\ref{sec3b}. Note that
 we use the MCMC code CosmoMC~\cite{Lewis:2002ah,CosmoMCsite}
 in doing this. The cosmographic formulae for $D_L$ and $\cal K$
 are given in Eqs.~(\ref{eq15}) and (\ref{eq17}) respectively,
 while one can calculate $D_V(z)$ by using $D_L$ given
 in Eq.~(\ref{eq15}),
 and $z/h(z)=[\,z/(1+z)\,]\cdot [\,dD_L/dz-D_L/(1+z)\,]$
 from Eq.~(\ref{eq16}).

As mentioned above, the Hubble constant $H_0$ (which is also
 the first cosmographic parameter) has not been involved in
 the time delay data from GRBs and the $D_V(0.35)/D_V(0.2)$
 data from BAO, while it is a nuisance parameter in the JLA
 SNIa dataset and can be  marginalized. On the other hand, if
 one consider the cosmographic formulae only up to 2nd order,
 the corresponding Taylor series cannot be enough general to
 include many cosmological models as its special cases, and
 hence the conclusions become not so robust. However, if one
 consider the cosmographic formulae up to very high order, too
 many cosmographic parameters will be involved and hence the
 constraints become very loose. On balance, here we consider
 the cosmographic formulae up to 3rd and 4th orders one by one.

In the case of 3rd order cosmography with respect to redshift
 $z$ (labeled as ``\,$z-j$\,''), namely we only consider the
 Taylor series for $D_L(z)$, ${\cal K}(z)$, $D_V(z)$ ... up to
 3rd order and ignore all the higher order terms
 ${\cal O}(z^4)$, there are only two free cosmographic
 parameters $q_0$ and $j_0$ besides the nuisance parameter
 $H_0$. So, the free parameters under consideration are
 $\left\{q_0,\, j_0,\, a_{LIV},\, b\right\}$. The total
 $\chi^2$ is given by
 $\chi^2_{tot}=\chi^2_{GRB}+\chi^2_{JLA}+\chi^2_{BAO}$. By
 fitting the cosmographic formulae to the combined GRB+JLA+BAO
 observational data, we obtain the $1\sigma$ and $2\sigma$
 constraints on the LIV parameters ($a_{LIV}$,~$b$) and the
 cosmographic parameters ($q_0$,~$j_0$), which are presented
 in the 2nd column of Table~\ref{tab1}. We also present the
 2D marginalized $1\sigma$ and $2\sigma$ contours and the 1D
 distributions of the LIV parameters and the cosmographic
 parameters in Fig.~\ref{fig1}. From Fig.~\ref{fig1} and the
 2nd column of Table~\ref{tab1}, we find a fairly weak hint
 for LIV with a non-zero $a_{LIV}$ (slightly beyond $1\sigma$
 confidence region), while $a_{LIV}=0$ is still consistent
 with the observational data within $2\sigma$ confidence
 region.

In the case of 4th order cosmography with respect to redshift
 $z$ (labeled as ``\,$z-s$\,''), namely we only consider the
 Taylor series for $D_L(z)$, ${\cal K}(z)$, $D_V(z)$ ... up to
 4th order and ignore all the higher order terms
 ${\cal O}(z^5)$, there are three free cosmographic parameters
 $q_0$, $j_0$ and $s_0$ besides the nuisance parameter $H_0$.
 So, the free parameters under consideration are
 $\left\{q_0,\, j_0,\, s_0,\, a_{LIV},\, b\right\}$. By
 fitting the cosmographic formulae to the combined GRB+JLA+BAO
 observational data, we obtain the $1\sigma$ and $2\sigma$
 constraints on the LIV parameters ($a_{LIV}$,~$b$) and the
 cosmographic parameters ($q_0$, $j_0$, $s_0$), which are
 presented in the 3rd column of Table~\ref{tab1}. We also
 present the 2D marginalized $1\sigma$ and $2\sigma$ contours
 and the 1D distributions of the LIV parameters and the
 cosmographic parameters in Fig.~\ref{fig2}. From
 Fig.~\ref{fig2} and the 3rd column of Table~\ref{tab1}, we
 find again a weak hint for LIV with a non-zero $a_{LIV}$
 (beyond $1\sigma$ confidence region), while $a_{LIV}=0$ is
 still consistent with the observational data within $2\sigma$
 confidence region.

So, in both cases of cosmography with respect to redshift $z$,
 LIV with a non-zero $a_{LIV}$ is slightly favored by the
 observational data. On the other hand, from Figs.~\ref{fig1},
 \ref{fig2}, and Table~\ref{tab1}, it is easy to see that in
 both cases, the deceleration parameter $q_0$ is negative
 beyond $2\sigma$ confidence region, and the jerk $j_0$ is
 positive also beyond $2\sigma$ confidence region. Thus,
 from the definitions in Eq.~(\ref{eq12}), this means that
 an accelerating universe ($q_0<0$) is strongly favored, while
 the acceleration is still increasing ($j_0>0$).


\begin{table}[tb]
 \renewcommand{\arraystretch}{2.0}
 \begin{center}
 \vspace{-4mm}  
 \begin{tabular}{c|c|c} \hline\hline
\ Parameters \ &  Case $\,z-j$  &  Case $\,z-s$  \\ \hline
$a_{LIV}$ & \ $-0.0056134_{-0.0033430}^{+0.0032712}\,(1\sigma)\,_{-0.0074206}^{+0.0073291}\,(2\sigma)$ \ & \ $-0.0358845_{-0.0180408}^{+0.0236884}\,(1\sigma)\,_{-0.0385662}^{+0.0367396}\,(2\sigma)$  \  \\  \hline
$b$ & \ $-0.0009220_{-0.0019603}^{+0.0018053}\,(1\sigma)\,_{-0.0038621}^{+0.0041510}\,(2\sigma)$ \ & \ $0.0153007_{-0.0135162}^{+0.0098136}\,(1\sigma)\,_{-0.0202124}^{+0.0214681}\,(2\sigma)$  \  \\  \hline
$q_0$ & \ $-0.4516519_{-0.0799675}^{+0.0802934}\,(1\sigma)\,_{-0.1641987}^{+0.1530627}\,(2\sigma)$ \ & \ $-0.5464667_{-0.0431929}^{+0.0501635}\,(1\sigma)\,_{-0.0981299}^{+0.0958649}\,(2\sigma)$  \  \\  \hline
$j_0$ & \ $0.5062922_{-0.1669237}^{+0.1333122}\,(1\sigma)\,_{-0.3016949}^{+0.3155159}\,(2\sigma)$ \ & \ $1.1111259_{-0.2501652}^{+0.1301341}\,(1\sigma)\,_{-0.3991027}^{+0.5109921}\,(2\sigma)$  \  \\   \hline
$s_0$ &   & \ $-0.2812477_{-0.5722327}^{+0.3564938}\,(1\sigma)\,_{-0.9656683}^{+1.0999906}\,(2\sigma)$  \  \\   \hline
 \hline
 \end{tabular}
 \end{center}
 \vspace{-2mm}  
 \caption{\label{tab1} The mean with $1\sigma$ and $2\sigma$
 uncertainties of the LIV parameters and the cosmographic
 parameters for the cases of 3rd order (labeled
 as ``\,$z-j$\,'', the 2nd column) and 4th order (labeled as
 ``\,$z-s$\,'', the 3rd column) cosmography with respect to
 redshift $z$, respectively. Note that $a_{LIV}$ and $b$ are
 given in units of seconds, while $q_0$, $j_0$, $s_0$ are all
 dimensionless. See the text for details.}
 \end{table}



\section{Constraints on LIV via the cosmographic
 approach with~respect~to $y$-shift}\label{sec4}


\subsection{Cosmographic approach with respect to
 $y$-shift}\label{sec4a}

It is easy to see that the key of cosmography is to expand
 the quantities under consideration as a Taylor series. In the
 original version of cosmography, the relevant quantities are
 expanded with respect to redshift $z$. However, it is well
 known that such a Taylor series converges only for small $z$
 around $0$, and it might diverge at high redshift $z>1$.
 A possible remedy is to replace $z$ with the so-called
 $y$-shift, $y\equiv z/(1+z)$ (see e.g.~\cite{Cattoen:2008th,
 Cattoen:2007id,Vitagliano:2009et,Xia:2011iv,Zhou:2016nik}).
 In this case, $y<1$ holds in the whole cosmic
 past $0\leq z<\infty$, and hence the Taylor series with
 respect to $y$-shift converges. So, here we also consider the
 cosmographic approach with respect to $y$-shift, $y\equiv z/(1+z)$.

We can expand the dimensionless luminosity distance $D_L$
 (equivalent to $d_L$) in terms of a Taylor series with respect
 to $y$ (see e.g.~\cite{Bamba:2012cp,Vitagliano:2009et,
 Zhou:2016nik} for details),
 \bea
 D_L(y) &=& y+ \frac{1}{2}\left(3-q_0\right)y^2 +
 \frac{1}{6}\left(11-5q_0+3q_0^2-j_0\right)y^3 \nonumber\\[1mm]
  & &+\frac{1}{24}\left(50-26q_0+21q_0^2-15q_0^3-7j_0+10q_0j_0
  +s_0\right) y^4+{\cal O}\left(y^5\right)\,.\label{eq26}
 \eea
 Alternatively, one can derive Eq.~(\ref{eq26}) by substituting
 $z=y/(1-y)=y+y^2+y^3+y^4+{\cal O}(y^5)$ into Eq.~(\ref{eq15})
 and then rearranging it as a series in terms
 of $y$. Similarly, we can also substitute
 $z=y/(1-y)=y+y^2+y^3+y^4+{\cal O}(y^5)$ into
 Eq.~(\ref{eq17}) and then rearrange it as a series in terms of $y$,
 \bea
 {\cal K}(y) &=& y+\left(1-\frac{q_0}{2}\right)y^2+\left(
 1-\frac{2}{3}q_0+\frac{q_0^2}{2}-\frac{j_0}{6}\right)y^3
 \nonumber \\[1mm]
 & &+\left(1-\frac{3}{4}q_0+\frac{3}{4}q_0^2-\frac{5}{8}q_0^3
 -\frac{j_0}{4}+\frac{5}{12}q_0 j_0+\frac{s_0}{24}\right)y^4
 +{\cal O}\left(y^5\right)\,.\label{eq27}
 \eea
 Of course, one can also expand $D_V$ as a Taylor series in terms of
 $y$ by substituting $z=y/(1-y)=y+y^2+y^3+y^4+{\cal O}(y^5)$
 into Eq.~(\ref{eq24}) and then rearranging it as a series in
 terms of $y$. Note that Eq.~(\ref{eq16}) is still useful
 in doing this.


\begin{table}[tb]
 \renewcommand{\arraystretch}{2.0}
 \begin{center}
 \vspace{-4mm}  
 \begin{tabular}{c|c|c} \hline\hline
\ Parameters \ &  Case $\,y-j$  &  Case $\,y-s$  \\ \hline
$a_{LIV}$ & \ $-0.0839179_{-0.0347786}^{+0.0386001}\,(1\sigma)\,_{-0.0695886}^{+0.0692573}\,(2\sigma)$ \ & \ $0.0045945_{-0.0126229}^{+0.0150052}\,(1\sigma)\,_{-0.0296695}^{+0.0290441}\,(2\sigma)$  \  \\  \hline
$b$ & \ $0.0384035_{-0.0190885}^{+0.0176107}\,(1\sigma)\,_{-0.0346907}^{+0.0349654}\,(2\sigma)$ \ & \ $-0.0063665_{-0.0082512}^{+0.0070200}\,(1\sigma)\,_{-0.0158987}^{+0.0163490}\,(2\sigma)$  \  \\  \hline
$q_0$ & \ $-0.6736552_{-0.3089059}^{+0.2455526}\,(1\sigma)\,_{-0.5047697}^{+0.5268171}\,(2\sigma)$ \ & \ $-0.8314270_{-0.4659630}^{+0.2482399}\,(1\sigma)\,_{-0.6406471}^{+0.7825488}\,(2\sigma)$  \  \\  \hline
$j_0$ & \ $2.5771079_{-3.3925562}^{+3.4483891}\,(1\sigma)\,_{-6.2831097}^{+6.1780295}\,(2\sigma)$ \ & \ $8.7896032_{-3.7547503}^{+10.6351976}\,(1\sigma)\,_{-13.7152157}^{+11.2103968}\,(2\sigma)$  \  \\   \hline
$s_0$ &   & \ $136.7073059_{-132.2172699}^{+118.7269897}\,(1\sigma)\,_{-189.9219513}^{+195.5156860}\,(2\sigma)$  \  \\   \hline
 \hline
 \end{tabular}
 \end{center}
 \vspace{-2mm}  
 \caption{\label{tab2} The mean with $1\sigma$ and $2\sigma$
 uncertainties of the LIV parameters and the cosmographic
 parameters for the cases of 3rd order (labeled
 as ``\,$y-j$\,'', the 2nd column) and 4th order (labeled as
 ``\,$y-s$\,'', the 3rd column) cosmography with respect to
 $y$-shift $y=z/(1+z)$, respectively. Note that $a_{LIV}$ and
 $b$ are given in units of seconds, while $q_0$, $j_0$,
 $s_0$ are all dimensionless. See the text for details.}
 \end{table}



 \begin{center}
 \begin{figure}[tbp]
 \centering
 \vspace{-12mm}  
 \includegraphics[width=1.0\textwidth]{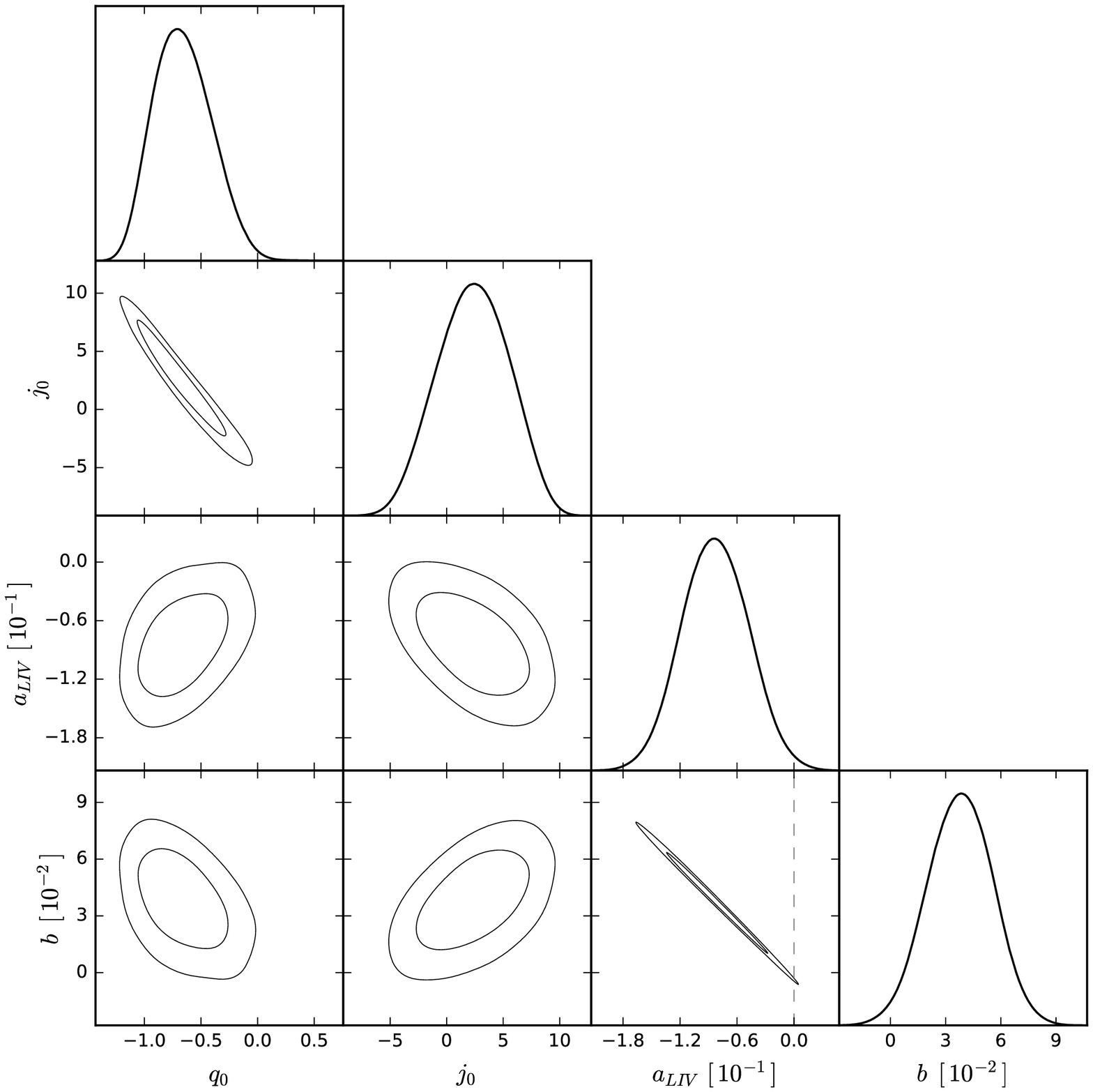}
 \caption{\label{fig3} The 2D marginalized $1\sigma$ and
 $2\sigma$ contours and the 1D distributions of the LIV
 parameters and the cosmographic parameters for the case of 3rd
 order cosmography with respect to $y$-shift (labeled as
 ``\,$y-j$\,''). Note that $a_{LIV}=0$ is also indicated by a
 dashed line in the $a_{LIV}-b$ panel, and $a_{LIV}$, $b$ are
 given in units of $10^{-1}$, $10^{-2}$, respectively. See
 the text and Table~\ref{tab2} for details.}
 \end{figure}
 \end{center}



 \begin{center}
 \begin{figure}[tbp]
 \centering
 \vspace{-12mm}  
 \includegraphics[width=1.0\textwidth]{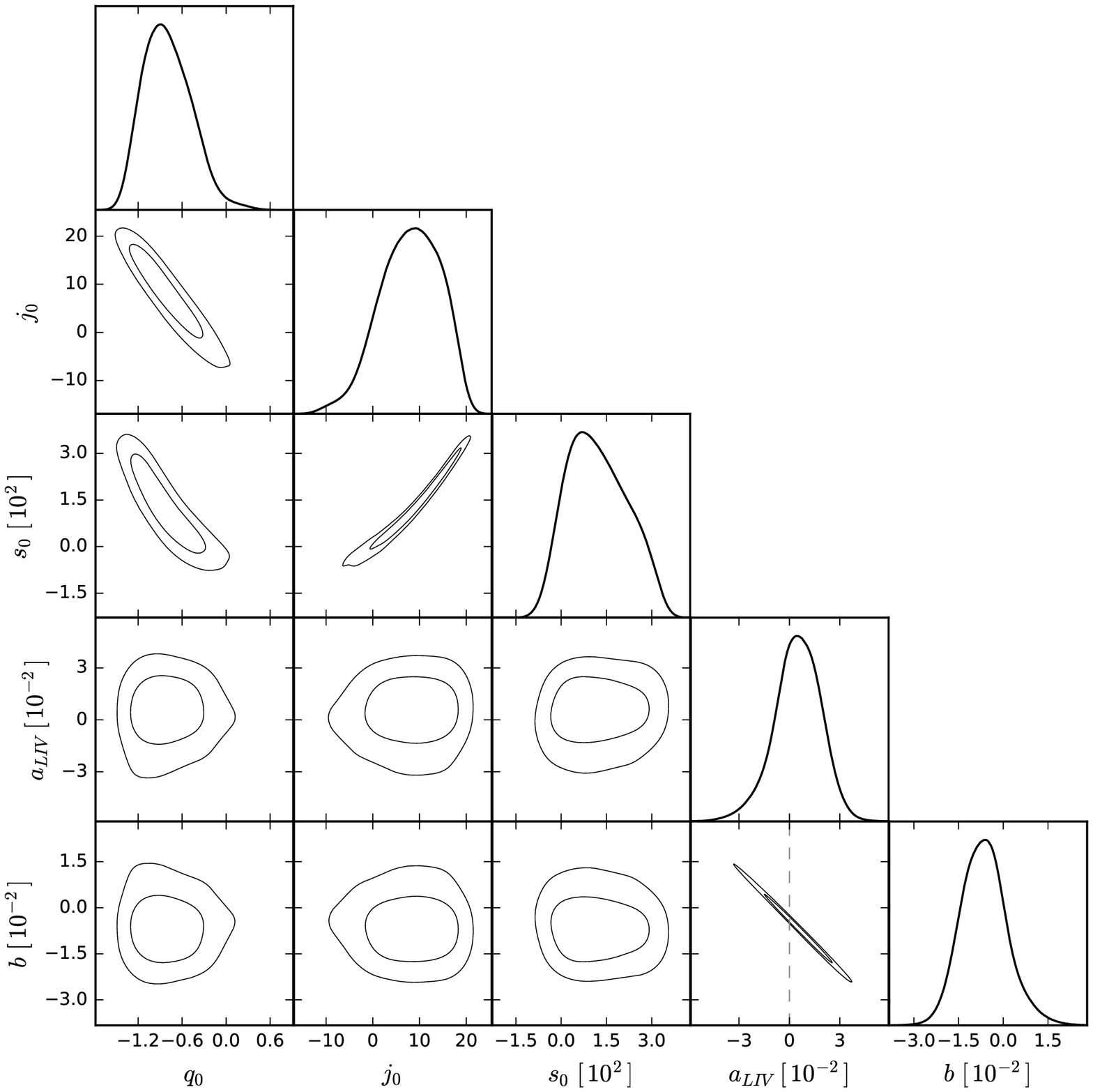}
 \caption{\label{fig4} The 2D marginalized $1\sigma$ and
 $2\sigma$ contours and the 1D distributions of the LIV
 parameters and the cosmographic parameters for the case of 4th
 order cosmography with respect to $y$-shift (labeled as
 ``\,$y-s$\,''). Note that $a_{LIV}=0$ is also indicated by a
 dashed line in the $a_{LIV}-b$ panel, and $s_0$, $a_{LIV}$,
 $b$ are given in units of $10^2$, $10^{-2}$, $10^{-2}$,
 respectively. See the text and Table~\ref{tab2} for details.}
 \end{figure}
 \end{center}


\vspace{-21mm}  


\subsection{Constraints on LIV}\label{sec4b}

Again, we constrain the LIV parameters $a_{LIV}$ and $b$,
 together with the cosmographic parameters, by using the
 observational data mentioned in Sec.~\ref{sec3b}. This is
 similar to Sec.~\ref{sec3c} actually, but the cosmographic
 formulae for $D_L$, $\cal K$, $D_V$ ... should instead use
 the ones with respect to $y$-shift $y=z/(1+z)$ given in
 Sec.~\ref{sec4a}.

In the case of 3rd order cosmography with respect to $y$-shift
 $y=z/(1+z)$ (labeled as ``\,$y-j$\,''), namely we only
 consider the Taylor series for $D_L(y)$, ${\cal K}(y)$,
 $D_V(y)$ ... up to 3rd order and ignore all the higher order
 terms ${\cal O}(y^4)$, there are only two free cosmographic
 parameters $q_0$ and $j_0$ besides the nuisance parameter
 $H_0$. So, the free parameters under consideration are
 $\left\{q_0,\, j_0,\, a_{LIV},\, b\right\}$. The total
 $\chi^2$ is given by
 $\chi^2_{tot}=\chi^2_{GRB}+\chi^2_{JLA}+\chi^2_{BAO}$. By
 fitting the cosmographic formulae to the combined GRB+JLA+BAO
 observational data, we obtain the $1\sigma$ and $2\sigma$
 constraints on the LIV parameters ($a_{LIV}$,~$b$) and the
 cosmographic parameters ($q_0$,~$j_0$), which are presented
 in the 2nd column of Table~\ref{tab2}. We also present the
 2D marginalized $1\sigma$ and $2\sigma$ contours and the 1D
 distributions of the LIV parameters and the cosmographic
 parameters in Fig.~\ref{fig3}. From Fig.~\ref{fig3} and the
 2nd column of Table~\ref{tab2}, we find a notable hint for
 LIV with a non-zero $a_{LIV}$ (around $2\sigma$ confidence
 region).

In the case of 4th order cosmography with respect to $y$-shift
 $y=z/(1+z)$ (labeled as ``\,$y-s$\,''), namely we only
 consider the Taylor series for $D_L(y)$, ${\cal K}(y)$,
 $D_V(y)$ ... up to 4th order and ignore all the higher order
 terms ${\cal O}(y^5)$, there are three free cosmographic
 parameters $q_0$, $j_0$ and $s_0$ besides the nuisance
 parameter $H_0$. So, the free parameters under consideration
 are $\left\{q_0,\, j_0,\, s_0,\, a_{LIV},\, b\right\}$. By
 fitting the cosmographic formulae to the combined GRB+JLA+BAO
 observational data, we obtain the $1\sigma$ and $2\sigma$
 constraints on the LIV parameters ($a_{LIV}$,~$b$) and the
 cosmographic parameters ($q_0$,\,$j_0$,\,$s_0$), which are
 presented in the 3rd column of Table~\ref{tab2}. We also
 present the 2D marginalized $1\sigma$ and $2\sigma$ contours
 and the 1D distributions of the LIV parameters and the
 cosmographic parameters in Fig.~\ref{fig4}. From
 Fig.~\ref{fig4} and the 3rd column of Table~\ref{tab2},
 we find that $a_{LIV}=0$ is fully consistent with the
 observational data, namely there is no evidence for LIV.

Although the results about LIV are quite different in both the
 cases of cosmography with respect to $y$-shift $y=z/(1+z)$,
 from Figs.~\ref{fig3}, \ref{fig4}, and Table~\ref{tab2}, it
 is easy to see that in both cases, the deceleration parameter
 $q_0$ is negative beyond $2\sigma$ confidence region. Thus,
 from the definition in Eq.~(\ref{eq12}), this means that
 an accelerating universe ($q_0<0$) is strongly favored.


 \begin{table}[tb]
 \renewcommand{\arraystretch}{1.5}
 \begin{center}
 \vspace{-4mm}  
 \begin{tabular}{lllll} \hline\hline
 \ Cosmography \quad\quad &  $z-j$ &  $z-s$  & $y-j$  & $y-s$  \\ \hline
 \ $\chi^2_{min}$   &  871.3198 \quad\quad & 860.0310 \quad\quad & 868.3670 \quad\quad & 871.9442 \ \\
 \ $\kappa$   &  4 &  5 &  4 & 5 \ \\
 \ $\chi^2_{min}/dof$   &  1.1287 & 1.1155 & 1.1248 & 1.1309 \ \\
 \ $ \Delta {\rm BIC} $   & 4.6346 & 0 & 1.6818 & 11.9132 \ \\
 \ $ \Delta {\rm AIC} $   & 9.2888 & 0 & 6.3360 & 11.9132 \ \\
 \ Rank  &  3  &  1  &  2  &  4  \ \\
 \hline\hline
 \end{tabular}
 \end{center}
 \vspace{-2mm}  
 \caption{\label{tab3} Comparing the four cases of cosmography
 considered in the present work, namely 3rd order (labeled as
 ``\,$z-j$\,'') and 4th order (labeled as ``\,$z-s$\,'')
 cosmography with respect to redshift $z$, as well as 3rd order
 (labeled as ``\,$y-j$\,'') and 4th order (labeled as ``\,$y-s$\,'')
 cosmography with respect to $y$-shift $y=z/(1+z)$. See the
 text for details.}
 \end{table}



\section{Concluding remarks}\label{sec5}

Since Lorentz invariance plays an important role in modern
 physics, it is of interest to test the possible LIV. The
 time-lag (the arrival time delay between light curves in
 different energy bands) of GRBs has been extensively used
 to this end. However, to our best knowledge, one or more
 particular cosmological models were assumed {\it a~priori}
 in (almost) all of the relevant works in the literature.
 So, this makes the results on LIV in those works
 model-dependent and hence not so robust in fact. In the
 present work, we try to avoid this problem by using a
 model-independent approach. We calculate the time delay
 induced by LIV with the cosmic expansion history given in
 terms of cosmography, without assuming any particular
 cosmological model. Then, we constrain the possible LIV with
 the observational data from GRBs, SNIa and BAO, and find weak
 hints for LIV with non-zero $a_{LIV}$ in 3 of 4 cases of
 cosmography considered in the present work.

It is of interest to compare the 4 cases of cosmography
 considered here. As mentioned above, they are labeled as
 ``\,$z-j$\,'', ``\,$z-s$\,'', ``\,$y-j$\,'' and ``\,$y-s$\,'',
 respectively. Since they have different free parameters
 and the correlations between model parameters are fairly
 different, it is not suitable to directly compare their
 confidence level contours. Instead, it is more appropriate to
 compare them from the viewpoint of goodness-of-fit. A
 conventional criterion for model comparison in the literature
 is $\chi^2_{min}/dof$, in which the degree of freedom
 $dof={\cal N}-\kappa$, while $\cal N$ and $\kappa$ are the
 number of data points and the number of free model parameters,
 respectively. On the other hand, there are other criteria
 for model comparison in the literature. The most
 sophisticated criterion is the Bayesian evidence (see
 e.g.~\cite{Liddle:2007fy} and references therein). However,
 the computation of Bayesian evidence usually consumes a large
 amount of time and power. As an alternative, one can consider
 some approximations of Bayesian evidence, such as the
 so-called Bayesian Information Criterion (BIC) and Akaike
 Information Criterion (AIC). The BIC is defined
 by~\cite{Schwarz:1978}
 \be{eq28}
 {\rm BIC}=-2\ln{\cal L}_{max}+\kappa\ln {\cal N}\,,
 \ee
 where ${\cal L}_{max}$ is the maximum likelihood. In the
 Gaussian cases, $\chi^2_{min}=-2\ln{\cal L}_{max}$. So, the
 difference in BIC between two models is given by
 $\Delta{\rm BIC}=\Delta\chi^2_{min}+\Delta\kappa\ln {\cal N}$.
 The AIC is defined by~\cite{Akaike:1974}
 \be{eq29}
 {\rm AIC}=-2\ln{\cal L}_{max}+2\kappa\,.
 \ee
 The difference in AIC between two models is
 given by $\Delta{\rm AIC}=\Delta\chi^2_{min}+2\Delta\kappa$.
 In Table~\ref{tab3}, we present
 $\chi^2_{min}/dof$, $\Delta$BIC and $\Delta$AIC for the 4
 cases of cosmography considered in this work. Note that
 ``\,$z-s$\,'' has been chosen to be the fiducial model when
 we calculate $\Delta$BIC and $\Delta$AIC. Clearly, from the
 viewpoint of all the three criteria $\chi^2_{min}/dof$, BIC
 and AIC, ``\,$z-s$\,'' is the best, and ``\,$y-s$\,'' is the
 worst. Together with the fact that we find no evidence for
 LIV only in the case ``\,$y-s$\,'' while there are weak hints
 for LIV in the other three cases ``\,$z-s$\,'', ``\,$y-j$\,''
 and ``\,$z-j$\,'', we conclude that LIV with non-zero
 $a_{LIV}$ is slightly favored by the observational data.

In the literature, there exist many relevant works on the
 possible LIV. In addition to the weak evidence for LIV found
 in~\cite{Ellis:2005wr} (its Erratum should be seriously
 considered), some further works (e.g.~\cite{Biesiada:2009zz,
 Xu:2016zxi,Xu:2016zsa,Wei:2016exb}) supported this conclusion
 of~\cite{Ellis:2005wr}. In particular, a strong evidence for
 LIV was claimed in~\cite{Xu:2016zxi,Xu:2016zsa}. Our results
 obtained in the present work could be regarded as a new
 support. On the other hand, one should also be aware of the
 contrary claim that there is no evidence for LIV (see
 e.g.~\cite{Pan:2015cqa}). Besides, most of the relevant works
 in the literature kept silence and just put a lower bound on
 the possible LIV energy scale $E_{QG}$. Therefore, the debate
 on LIV has not been settled so far. More and
 better observational data from e.g.~GRBs
 are needed. New ideas to test LIV are also desirable.

In the present work, we consider the cosmographic approaches
 with respect to redshift $z$ and $y$-shift $y=z/(1+z)$. The
 first one might diverge at high redshift $z>1$, while the
 second one can alleviate this problem since $y<1$ in the
 whole cosmic past $0\leq z<\infty$, and hence the Taylor
 series with respect to $y$ converges. However, there still
 exist several serious problems in the case of $y=z/(1+z)$.
 The first is that the error of a Taylor approximation
 throwing away the higher order terms will become unacceptably
 large when $y$ is close to~$1$ (say, when $z>9$). The second
 is that the cosmography in terms of $y=z/(1+z)$ cannot work
 well in the cosmic future $-1<z<0$. The Taylor series with
 respect to $y=z/(1+z)$ does not converge when $y<-1$ (namely
 $z<-1/2$), and it drastically diverges when $z\to -1$ (it is
 easy to see that $y\to -\infty$ in this case). So, the
 $y$-shift cosmography fails to predict the future evolution
 of the universe. In~\cite{Zhou:2016nik}, two new
 generalizations of cosmography inspired by the Pad\'e
 approximant have been proposed, which can avoid or at least
 alleviate the problems of ordinary cosmography mentioned
 above. The model-independent constraints on LIV via these
 two new cosmographic approaches proposed
 in~\cite{Zhou:2016nik} deserve further investigation.

In most of the relevant works on LIV (including the present
 work), the intrinsic time-lag of GRBs is actually oversimplified by
 assuming $\Delta t_{\rm int}=b\left(1+z\right)$. Noting that
 the factor $(1+z)$ comes from the cosmic expansion, this
 assumption means that all GRBs have the same intrinsic time
 delay (characterized by the constant parameter $b$) in the
 source frame. In e.g.~\cite{Wei:2016exb}, an energy-dependent
 time-lag was proposed. We consider that a more realistic assumption
 for the intrinsic time-lag of GRBs is important to robustly
 constrain the possible LIV in the future work. Deeper
 understanding on the observed time-lag of GRBs is also
 desirable (nb.~Eq.~(\ref{eq6})), especially the ones other
 than LIV-induced time delay.

As mentioned in Sec.~\ref{sec1}, the possible LIV is commonly
 accompanied with a non-trivial dispersion relation. A signal
 of energy $E$ that travels a distance $L$ acquires a time
 delay (measured with respect to the ordinary case of an
 energy-independent speed $c$),
 namely $\Delta t\sim\xi (E/E_{QG})(L/c)$. Although the
 QG effect is expected to be very weak, a very long distance
 $L$ can still make it testable. This point can be clearly seen
 from Eq.~(\ref{eq5}), namely a large time delay
 $\Delta t_{LIV}$ follows a high redshift $z$. GRBs are among
 the most powerful sources in the universe. Their high energy
 photons in the gamma-ray band are almost immune to dust
 extinction, and hence they have been observed up to redshift
 $z\sim 8-9$~\cite{Salvaterra:2009,Cucchiara:2011hd}, while the
 maximum redshift of GRBs is expected to be $10$ or even
 larger~\cite{Bromm:2002xt,Lin:2003ff}. Therefore, GRBs at high
 redshift can be used to test the possible LIV which induces a
 large time delay. In the present work, we find weak hints for
 LIV by using the time delay dataset from 35 GRBs with
 redshifts up to $z=6.29$~\cite{Ellis:2005wr} via
 the cosmographic approach. Our results suggest that the
 possible LIV should be taken seriously. The physical mechanism
 for LIV might be the spacetime foam predicted in most of the
 quantum gravity theories (e.g. string theory, loop quantum
 gravity, and doubly special relativity). In addition, we would
 like to mention the so-called Standard-Model Extension
 (SME)~\cite{Mewesworks}, which also provides a field theory
 framework for LIV. The deep physics behind LIV deserves
 serious consideration.


\section*{ACKNOWLEDGEMENTS}
We thank the anonymous referee for quite useful comments
 and suggestions, which helped us to improve this work.
 We are grateful to Profs.~Lixin~Xu, Zong-Kuan~Guo, Yu~Pan
 and Weiqiang~Yang for helpful discussions. We also thank
 Ya-Nan~Zhou, Jing~Liu, Zu-Cheng~Chen, Shou-Long~Li, Hong-Yu~Li
 and Dong-Ze~Xue for kind help and discussions. This work
 was supported in part by NSFC under Grants
 No.~11575022 and No.~11175016.

\renewcommand{\baselinestretch}{1.0}


\end{document}